\documentclass{PoS}

\title{Spectral functions at non-zero momentum in hot QCD}

\ShortTitle{Spectral functions at non-zero momentum in hot QCD}

\author{Gert Aarts, Chris Allton, \speaker{Justin Foley}, Simon Hands\\

        Department of Physics, Swansea University, Swansea, United Kingdom\\

        E-mail: \email{g.aarts@swansea.ac.uk},

        \email{c.allton@swansea.ac.uk},

        \email{j.foley@swansea.ac.uk}, 
	
        \email{s.hands@swansea.ac.uk}}

\author{and Seyong Kim\\

        Department of Physics, Sejong University, Seoul, Korea\\

        E-mail: \email{skim@sejong.ac.kr}}

\abstract{We present results for meson spectral functions at non-zero 
momentum at temperatures both below and above $T_c$, obtained 
in quenched simulations for a number of valence quark masses. For the 
lightest quark masses, a clear difference between the spectral functions 
on cold and hot lattices is observed.}

\FullConference{XXIVth International Symposium on Lattice Field Theory\\

                July 23-28, 2006\\

                Tucson, Arizona, USA}

\begin{document}

\section{Introduction} 
The experimental program at RHIC has generated much interest 
in the properties of the quark-gluon plasma (QGP) in the theoretical 
community, not least amongst lattice gauge theorists. 
Many of the dynamical properties of the QGP can only be determined 
by studying the relevant spectral functions $\rho (\omega, \mathbf{p})$.  
These are related to the corresponding euclidean-time 
correlators via an integral equation 
\begin{eqnarray}
G \left( \tau, \mathbf{p} \right) = \int_{0}^{\infty} 
\frac{d \omega} {2 \pi} K \left( \tau, \omega \right) \rho \left(\omega, \mathbf{p} \right), 
\hspace{1cm} K(\tau, \omega) = 
\frac{\cosh [ \omega  \left( \tau - 1/2T  \right) ]     } 
{ \sinh \left( \omega/2T \right) }.
\label{corr}
\end{eqnarray}
The survival of bound charmonium states in $N_{f}=2$ QCD well above the transition 
temperature~\cite{charmonium}, the calculation of dilepton and photon production 
rates~\cite{dilepton} and the calculation of QGP transport coefficients~\cite{gupta,nakamura} are all 
topics which have been studied on the lattice.  
Determining a spectral function from a correlator, evaluated at 
a finite number of points in a lattice simulation,
is an ill-posed problem.
However, progress can be made 
by using the maximum entropy method (MEM)~\cite{mem} to determine the 
most likely form for the spectral function given both the data 
and a set of prior assumptions for the spectral function.  
This prior information is encoded in a default model  
which is used in the MEM analysis, and one must be careful   
that the results obtained from this analysis 
are stable under reasonable variations 
of the default model.  

We are particularly interested in calculating the hydrodynamical contribution in QCD spectral functions.
Transport coefficients can be obtained from the slope of zero-momentum spectral functions at $\omega=0$,
for example, the electrical conductivity is given by  
\begin{eqnarray} 
\sigma =  \lim_{\omega \rightarrow 0} \frac{\rho_{em}\left(\omega, \mathbf{0}\right)} {6 \omega},
\end{eqnarray}
where $\rho_{em}$ is the spectral function for the conserved  
vector current.
Unfortunately, lattice correlators are extremely insensitive 
to the details of spectral functions at small frequencies~\cite{gertproc}.  
From general considerations, one knows that 
$\rho \left(\omega, \mathbf{p} \right) \sim \omega$ for $\omega \ll T$, 
and in the same regime the kernel can be written 
$
K\left(\tau, \omega \right) = (2T/\omega)
+ \mathcal{O}\left(\omega/T\right)
$.
Substituting these expressions into Eq.~(\ref{corr}),
one sees that the bulk of the contribution of the spectral function in the 
low frequency regime amounts to just a constant shift in the 
euclidean correlator. Therefore, one expects 
that the low-frequency part of a spectral function is 
the most difficult to compute using MEM. 

As a first step, we have undertaken a study of meson spectral 
functions above the transition temperature. These functions 
are obtained from the euclidean correlators  
$G(\tau, \mathbf{x}) = 
\langle J(\tau, \mathbf{x}) J^{\dagger} \left(0, \mathbf{0} \right) \rangle$ 
where $J(\tau, \mathbf{x})$ are quark field bilinears 
\begin{eqnarray}
J \left( \tau, \mathbf{x} \right) = \bar{\psi} (\tau, \mathbf{x}) \Gamma \psi( \tau, \mathbf{x}), \hspace{5mm}
\Gamma = 1, \gamma_{5}, \gamma_{i}.
\end{eqnarray}
We also computed these spectral functions for a number 
of values of spatial momentum.
Changing the momentum allows us to investigate whether momentum-dependent 
features can be found, which may be helpful in identifying non-trivial 
hydrodynamic structure. 
 
This work is particularly timely because there now 
exist predictions from AdS/CFT  
for finite-momentum spectral functions in $\mathcal{N}=4$ Super Yang-Mills theory~\cite{kovtun,teaney},   
and a comparison with lattice results is of immediate  
interest.

\section{Simulation Details} 
This study has been performed on sets of 100 quenched gauge configurations 
generated with a Wilson action on two lattices at temperatures both below 
and above $T_{c}$. 
The parameter values for these configurations are a subset of the values used 
in Ref.~\cite{bielefeld}. 
The `cold' configurations were generated on a $48^{3} \times 24$ lattice 
with $\beta = 6.5$. The temperature on this lattice was approximately $160~\rm{MeV}$. 
The second set of configurations were produced on a $64^{3} \times 24$ lattice  
with $\beta = 7.192$. In this case $ T \sim 420~\rm{MeV}$, which is about one and a half 
times the transition temperature.  
Staggered fermions were used for the valence quarks, and propagators were 
generated at three bare quark masses, $am = 0.01, 0.05, 0.125$. 
We are interested in comparing our results to hydrodynamic predictions 
and so we want to compute spectral functions for a number of momenta 
$p/T \lesssim 1$. However, for an isotropic lattice with periodic boundary conditions 
in the spatial directions, the lowest non-zero momentum in units of temperature is 
$p/T = 2 \pi N_{\tau}/N_{\sigma}$, where $N_{\tau}$ is the number of lattice sites in the 
temporal direction and $N_{\sigma}$ is the extent in the spatial directions.  
On the hot lattice, this corresponds to a lowest non-zero momentum 
$p/T \sim 2.4$.
To circumvent this problem and reach sufficiently small momenta, we 
use twisted boundary conditions for the valence quarks~\cite{flynn}.  
We computed propagators at four different twist angles and, 
using these together with Fourier transforms, we
are able to access $\sim20$ low-lying momenta. 

\subsection{$Z_{3}$ symmetry}
It is well known that the pure gauge action has a global $Z_{3}$ symmetry. 
It is left unchanged under the multiplication of the temporal links 
by an element of $Z_{3}$, the centre group of SU(3).  
Of course, this is a symmetry of the quenched theory only and is broken 
by the fermion determinant appearing in the partition function of 
full QCD.
The Polyakov line acts as an order parameter for the centre symmetry 
and the transition to the deconfined phase corresponds to
spontaneous breaking of the $Z_{3}$ symmetry. 
The $Z_{3}$ vacua can be distinguished by the Polyakov loop expectation value, 
which can be real or complex: $\langle P \rangle = 1, \exp\left(\pm i 2 \pi/3\right)$.
In Ref.~\cite{staggered} it was shown for staggered fermions that  
the nature of the chiral transition in the real vacuum is markedly different 
from that of the vacua with complex Polyakov loops. Above $T_{c}$ in the real phase  
the chiral condensate drops to zero with vanishing 
quark mass, signalling the restoration of chiral symmetry. 
This is akin to what one might expect in full QCD.       
In the calculation of the quark propagators we chose our link variables such that the phase 
of the Polyakov loop is real, and observe a similar signal of 
chiral symmetry restoration: above $T_{c}$ for the lightest 
quark mass $m/T = 0.24 $ the pseudoscalar and scalar correlators are 
seen to coincide.   

\subsection{MEM analysis}
For Kogut-Susskind fermions, the lattice correlator receives a 
contribution from an unwanted staggered partner. In this 
case the expression for the correlator in Eq.~(\ref{corr}) becomes
\begin{eqnarray}
G \left(\tau, \mathbf{p} \right) = 2 \int_{0}^{\infty} \frac{d \omega} { 2 \pi} 
K(\tau,\omega) [ \rho \left( \omega, \mathbf{p} \right) - (-1)^{\tau/a} \tilde{\rho} \left( \omega, \mathbf{p} \right) ]. 
\label{stg}
\end{eqnarray}
To compute $\rho \left( \omega, \mathbf{p} \right)$, we perform independent 
MEM analyses on odd and even time slices and add the results to obtain 
the desired spectral function. In doing this, one has to be careful that   
the correlators defined on alternate time slices yield sensible, positive 
semi-definite spectral functions that can be handled by MEM. 
The main disadvantage of using staggered fermions compared to other fermion 
formalisms is that only half the available time slices are 
used in each MEM analysis. 
All the results presented here have been obtained using the standard 
Bryan analysis, although we have checked this approach  against 
the classic and historic MEM methods. 
The default model initially used in our analysis is motivated by the 
expected behaviour of the spectral function in the continuum at asymptotically large 
frequencies $ \rho_{\rm{default}} \left( \omega \right) \sim  \omega^{2}$. 

\section{Results}
\begin{figure}
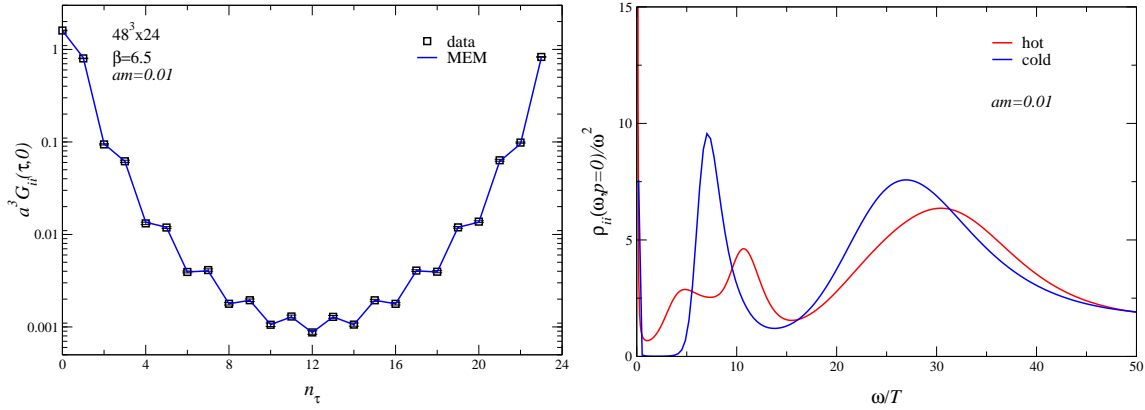

\resizebox{0.50\textwidth}{!}{
 \includegraphics{G_ViVi_m1_zaa.eps}
}
\resizebox{0.50\textwidth}{!}{
  \includegraphics{rho_ViVi_hotcold1.eps} 
}
  \caption[]{The plot on the left shows the  reconstructed correlator in the vector channel plotted along with the original Monte Carlo data points. The right-hand side shows zero-momentum spectral functions in the vector channel both above 
and below $T_{c}$.}
\label{tst}
\end{figure}

As a first check, we confirmed that it was possible 
to reconstruct the euclidean correlators from the results 
of the MEM analysis. The output for $\rho(\omega, \mathbf{p})$ 
and $\tilde \rho(\omega, \mathbf{p})$ were used to evaluate 
$G(\tau, \mathbf{p})$ by performing a simple numerical sum which approximated the 
integral in Eq.~(\ref{stg}).
An example of this 
is shown in the left-hand plot in Figure~\ref{tst}. 
This plots a reconstructed vector 
correlator, for the lightest quark mass, at vanishing spatial momentum below $T_{c}$, against the original numerical 
data points. The Monte Carlo results include 
a sum over the three vector polarisations.
The agreement between the original data and the reconstructed  
data points is excellent, and we find this to be true in general  
provided that the data is of high quality with sufficiently small 
error bars. As one might expect, when the data is noisier the match 
between the reconstructed correlator and the original data is less precise.

In the confined phase below $T_{c}$, the spectral functions 
are
peaked at energies which correspond to hadron resonances. 
Above $T_{c}$ the spectral functions look very different, particularly 
in the small frequency region. For example, the contribution from Landau damping 
means that these spectral functions are non-zero for all $\omega > 0$.
The right-hand side of Figure~\ref{tst} shows vector meson 
spectral functions at zero spatial momentum above and below $T_{c}$.
These have been rescaled by a factor of $\omega^{-2}$ to better reveal 
their detailed structure. 
In this plot, 
the melting of the hadronic state  
is evident. We also note that the spectral 
functions in both the hot and the cold phase show 
a large bump, centred at $\omega/T \sim 30$, 
which can attributed to lattice artifacts~\cite{b2,gert}. 

\begin{center}
\begin{figure}[ht]
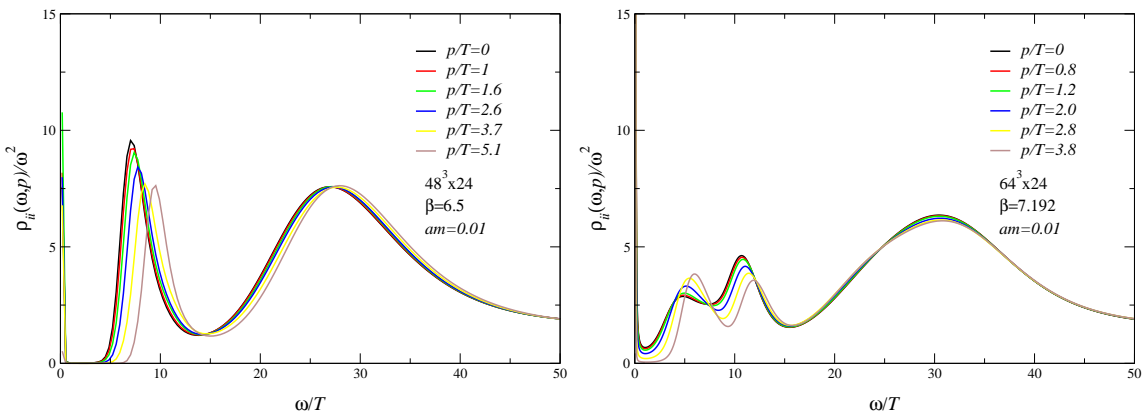

\resizebox{0.50\textwidth}{!}{
\includegraphics{rho_ViVi_mom6.eps}
}
\resizebox{0.50\textwidth}{!}{
\includegraphics{rho_ViVi_64x24_mom6.eps}
}
\caption[]{Vector spectral functions at the lightest quark mass for a range of momenta.}
\label{rhomomcold}
\end{figure}
\end{center}

Figure~\ref{rhomomcold} shows the vector 
spectral function for varying values 
of spatial momentum. Below $T_{c}$ the position 
of the spectral function peak shifts to higher energies with 
increasing spatial momentum, and the corresponding 
dispersion relation yields a speed of light which 
is close to unity.  
Note also that, although we are in the cold 
phase, the temporal extent of the lattice $( \sim 1.2~\rm{fm})$
is not large enough to allow us to compute the 
bound-state energies using
conventional lattice techniques, i.e. by 
fitting to the exponential decay of the correlator 
at sufficiently large times.  

\begin{figure}[ht]
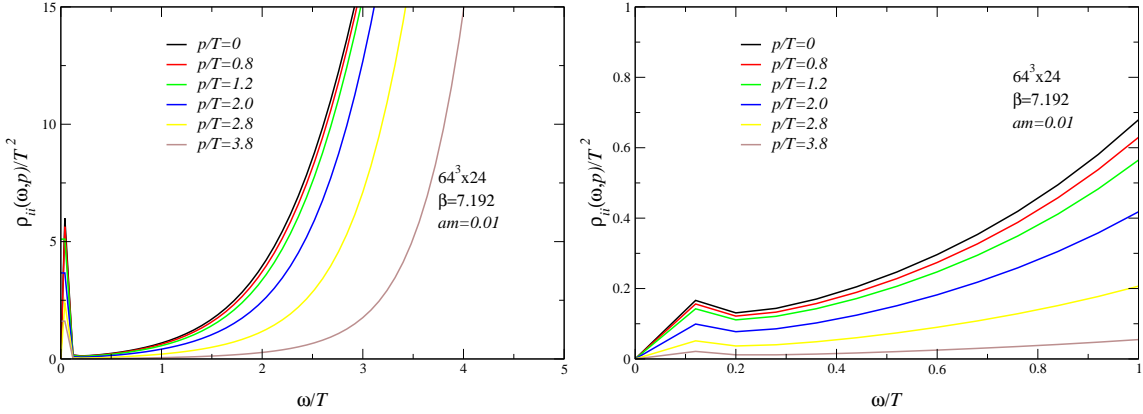

\resizebox{0.50\textwidth}{!}{
\includegraphics{rho_ViVi_mom6_T2.eps}
}
\resizebox{0.50\textwidth}{!}{
\includegraphics{rho_ViVi_mom6_T2_zoom2.eps}
}
\caption[]{Close-up of the spectral functions in the hot phase shown 
both with and without the unphysical peak at $\omega=0$.}
\label{nopeak}
\end{figure}

Above $T_{c}$, in the small $\omega$ region, there are very clear differences 
between spectral functions with different values of spatial momentum. 
In Figure~\ref{nopeak}, there appears to be a threshold energy which increases 
with increasing momentum.
The spectral functions are odd functions of $\omega$ and
the peak observed at 
$\omega \sim 0$ is an MEM artifact. 
Determining the precise origin 
of this artifact will require further analysis. For the purpose of this 
discussion, we
adopt a naive approach and simply disregard the unphysical peak.
The result of this is again shown in 
Figure~\ref{nopeak}. The spectral density has a bump at small values 
of $\omega$ which dies away with increasing momenta.  
This structure resembles the non-trivial structure 
expected from hydrodynamics and 
similar behaviour has been observed in an 
AdS/CFT calculation of spectral functions of the 
stress-energy tensor in $\mathcal{N}=4$ SYM~\cite{kovtun}.

\begin{figure}
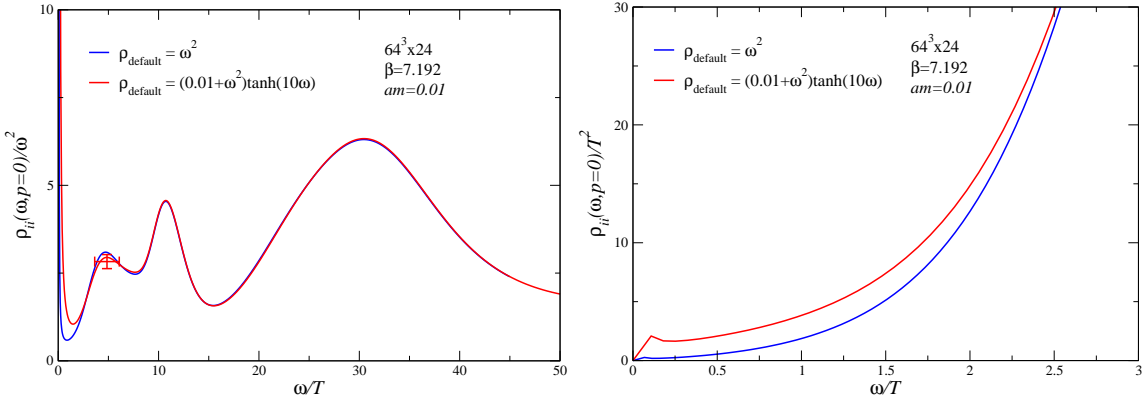

\resizebox{0.50\textwidth}{!}{
\includegraphics*{def2.eps}
}
\resizebox{0.50\textwidth}{!}{
\includegraphics*{def3.eps}
}
\caption[]{A comparison of zero-momentum vector spectral functions in the hot phase determined 
using different MEM default models as described in the text. The errorbar in the left-most plot gives an indication 
of the uncertainty in the MEM fits.}
\label{closeup1}
\end{figure}

 To test the reliability 
of these results, we repeated our analysis of the correlators 
with another default model. 
In this case we chose $\rho_{\rm{default}} \left( \omega \right) \sim \tanh\left(\omega/\delta\right) 
\left( \delta^{2} + \omega^{2} \right)$ so that for $\omega \ll \delta$ 
the default model varies linearly with $\omega$ rather than quadratically, and  
it again scales as $\omega^{2}$ for $\omega \gg \delta$.
For this analysis we took $\delta = 0.1$.  
A comparison of the zero-momentum vector spectral functions obtained using 
the original and new default model is shown on the left-hand plot in Figure~\ref{closeup1}. 
Here, one can see that for $ \omega/T \gtrsim 5$ the spectral functions 
are indistinguishable. Of course, there is no reason why they should differ, 
given that the two default models are virtually identical in this region.
However in the small $\omega$ region  
considerable differences appear. A close-up of this is also shown in 
Figure~\ref{closeup1}. Once again, we have removed the unphysical spike 
at the origin. We note that at low frequencies the spectral 
functions obtained using different default models do not 
agree quantitatively, although this difference may be consistent with the
uncertainty in the MEM fit indicated by the errorbar in the left-hand figure.
On the other hand, one might also argue  
that the original default model is completely at odds with the 
known behaviour of spectral functions for small $\omega$, and 
is not an appropriate default model for this regime.  
We also point out that, because of staggering in our correlators, 
we have used no more than twelve time slices in our analysis and 
it is possible that the results will converge as more time slices 
are used in the fit. 
Given the precise agreement between the reconstructed correlators and the Monte Carlo data  
shown earlier,
this discrepancy highlights the fact that 
lattice correlators are remarkably insensitive to the details 
of the spectral function at low frequencies. This result indicates  
how difficult a precise determination of transport coefficients 
will be.




\section{Conclusions/Outlook}
We have computed meson spectral functions 
at non-zero momentum using the maximum entropy method. 
Preliminary results have been obtained both in the cold confined phase 
and in the hot phase above $T_{c}$. Twisted boundary 
conditions have been employed to access a range of 
low-lying momenta in order to look for non-trivial 
hydrodynamic structure. The maximum entropy method 
can clearly distinguish between spectral functions 
with differing momenta. Qualitatively, the spectral 
functions show little variation under a change of default 
model. For example in the confined phase the positions of peaks which 
identify hadronic bound-states appear to be stable. 
However, a quantitative determination of the 
spectral functions at low frequencies is still lacking, and further 
work is required.   

The results presented here were obtained using 
staggered valence quark propagators. 
This is not the optimal choice for the MEM analysis.  
We are currently generating quark propagators using 
the non-perturbatively improved clover action~\cite{clover}.  
This will allow us to include all time slices of the resulting 
correlators in the MEM analysis, which we hope 
will benefit the analysis.

Ultimately, it may be that a much more sophisticated approach, 
including, for example, the use of highly anisotropic lattices~\cite{charmonium,dublin},
is required to precisely determine dynamical properties of QCD above 
$T_{c}$. However, the exploratory study described here certainly  
represents a progression towards this goal.


\begin{thebibliography}{99}
  \bibitem{charmonium} G. Aarts, C. R. Allton, R. Morrin, A. P. O Cais, M. B. Oktay, M. J. Peardon and J. I. Skullerud, hep-lat/0608009; PoS \textbf{LAT2005} 176 (2006) 
  \bibitem{dilepton} F. Karsch, E. Laermann, P. Petreczky, S. Stickan and I. Wetzorke, Phys. Lett. B \textbf{530} 147 (2002) 
  \bibitem{gupta} S. Gupta, Phys. Lett. B \textbf{597} 57 (2004)
  \bibitem{nakamura} A. Nakamura and S. Sakai, Phys. Rev. Lett. \textbf{94} 072305 (2005)
  \bibitem{mem} M. Asakawa, T. Hatsuda and Y. Nakahara, Prog. Part. Nucl. Phys. \textbf{46} 459 (2001)
  \bibitem{gertproc} G. Aarts and J. M. Martinez Resco, JHEP \textbf{0204} 053 (2002) 
  \bibitem{kovtun} P. Kovtun and A. Starinets, Phys. Rev. Lett. \textbf{96} 131601 (2006) 
  \bibitem{teaney} D. Teaney, Phys. Rev. D \textbf{74} 045025 (2006)
  \bibitem{bielefeld} S. Datta, F. Karsch, P. Petreczky and I. Wetzorke, Phys. Rev. D \textbf{69} 094507 (2004) 
  \bibitem{flynn} J. M. Flynn, A. J\"{u}ttner and C. T. Sachrajda, Phys. Lett. B \textbf{632} 313 (2006) 
  \bibitem{staggered} S. Chandrasekharan and N.H. Christ, Nucl. Phys. Proc. Suppl. \textbf{47} 527 (1996) 
  \bibitem{b2} F. Karsch, E. Laermann, P. Petreczky, S. Stickan, Phys. Rev. D \textbf{68} 014504 (2003) 
  \bibitem{gert} G. Aarts and J. M. Martinez Resco, Nucl. Phys. B \textbf{726} 93-108 (2005) 
  \bibitem{clover} M. L\"{u}scher, S. Sint, R. Sommer, P. Weisz and U. Wolff, Nucl. Phys. B \textbf{491} 323 (1997) 
  \bibitem{dublin} R. Morrin, A. P. O Cais, M. J. Peardon and S. M. Ryan, Phys. Rev. D \textbf{74} 014505 (2006) 
\end{thebibliography}
\end{document}